\documentclass[%
 reprint,
 amsmath,amssymb,
 aps,
]{revtex4-1}

\usepackage{graphicx}
\usepackage{dcolumn}
\usepackage{bm}
\usepackage{graphicx}
\usepackage{epstopdf}
\usepackage{float}
\usepackage{color}

\begin{document}
 \bibliographystyle{unsrt}
 \vfill

\title{Robust Quantum Control against Clock Noises in Multi-Qubit Systems}


\author{Hai-Jin Ding and Re-Bing Wu}%
\affiliation{Department of Automation, Tsinghua University, Beijing, 100084, China}
\email{rbwu@tsinghua.edu.cn}

\begin{abstract}
High-precision manipulation of multi-qubit quantum systems requires strictly clocked and synchronized multi-channel control signals. However, practical Arbitrary Waveform Generators (AWGs) always suffer from random signal jitters and channel latencies that induces non-ignorable state or gate operation errors. In this paper, we analyze the average gate error caused by clock noises, from which an estimation formula is derived for quantifying the control robustness against clock noises. This measure is then employed for finding robust controls via a homotopic optimization algorithm. We also introduce our recently proposed stochastic optimization algorithm, b-GRAPE, for training robust controls via randomly generated clock noise samples. Numerical simulations on a two-qubit example demonstrate that both algorithms can greatly improve the control robustness against clock noises. The homotopic algorithm converges much faster than the b-GRAPE algorithm, but the latter can achieve more robust controls against clock noises.
\end{abstract}


\maketitle

\section{\label{sec:level1}Introduction}
The multi-qubit manipulation is prevalent for quantum information processing with large-scale quantum circuits\cite{PhysRevA.29.1419,Nielsen2010,Preskill2018}. Towards fault-tolerant quantum computation, high precision control must be achieved above the error-correction threshold for the state and gate operations, and these have to be accomplished in presence of disturbances or uncertainties (e.g., pulse distortion, crosstalks and device noises \cite{Hincks2015}). Under such circumstances, robust single-shot controls are highly demanded so that the control precision is as insensitive as possible to the uncertainties or noises. Various algorithms have been proposed for this purpose, such as ensemble control for field inhomogeneity~\cite{PhysRevA.85.022302}, dynamical decoupling for environmental noises~\cite{DD}, STIRAP for control pulse shape errors~\cite{Vitanov2017}, DRAG algorithm for level leakages~\cite{PhysRevLett.103.110501}, detuning pulses~\cite{detuning,detuing2} for control field compensation, and sampling-based algorithms for generic parametric uncertainties in the Hamiltonian~\cite{Wu2018b}.

In this paper, we study robust controls against clock noises, a broadly existing but rarely considered noise source coming from clock signals in imperfect control signal generators in multi-qubit control systems. Ideally, the clock signals in different channels must be perfectly synchronized with the a common reference clock, but random timing errors always occur in realistic clock signals, especially in high-speed arbitrary waveform generators (AWG) \cite{Ryan2017,Ryan2017Hardware}. Typical clock noises include clock jitters (referred to as the timing deviation from a presumably periodic signal in relation to a reference clock signal) and channel latencies (referred to as the delay time from the AWG to the quantum chip to be controlled) \cite{Liu2017}. The clock noises randomly alter pulse areas and thus lead to stochastic operation errors.

In state-of-art AWG devices, the clock jitters can be managed to be at picosecond scale. The channel latencies vary from system to system, which are often much longer and much uncontrollable. For now, the affection of clock noises on the control precision are thought to be less important than other error sources (e.g., decoherence), but eventually they must be considered in extremely high-precision regime after other errors are mitigated~\cite{Ryan2017Hardware}. Therefore, it is deserved to study whether it is possible and how to find robust controls against clock noises.

To the authors' knowledge, clock noises have not been considered in the design of robust control of quantum gates. Related studies can be found in classical control systems (e.g., networked control systems with communication delays~\cite{Ding2018Distributed,CHIOU2018240}), in which the system's dynamics is described by an equivalent Markovian process driven by clock noises in the controller~\cite{Lamperski2016Optimal}. Stochastic optimal control theory can then be applied for state filtering and noise suppression ~\cite{Xu2002Calculation,Fontanelli2014A,Giorgi2015An}.
However, these results cannot be extended here to quantum control systems, because the required realtime feedback and communication between distributed agents are usually infeasible. Under most circumstances, one has to seek an open-loop (i.e., single-shot) robust control that is as insensitive as possible to the clock noises.

In this paper, we will first carry out a perturbation analysis on the average gate error induced by clock noises, from which an estimation formula is derived for evaluating the sensitivity of the control to clock noises. Then, two optimization algorithms are proposed for designing controls with enhanced robustness. The rest of the paper is arranged as follows. In Section \ref{Sec:error}, we analyze the average gate error caused by latency and jitter noises. In Section \ref{Sec:Three}, we propose a deterministic homotopic algorithm and a randomized b-GRAPE algorithm for the design of robust and high-precision controls, whose effectiveness are demonstrated in Section \ref{sec:Simulation} via numerical simulations on a two-qubit system. Finally, conclusions are drawn in Section \ref{Sec:Con}.

\section{Modeling and Analysis of clock-noise induced errors} \label{Sec:error}
\subsection{Quantum control in presence of clock noises}
Consider a quantum control system that involves $m$ control channels, and its unitary propagator is governed by $i\dot{U}(t)=H(t)U(t)$, where $U(t)\in \mathbb{C}^{N\times N}$ represents the quantum gate operation starting from $U(0)=\mathbb{I}$. The controlled Hamiltonian reads
\begin{equation}
H(t)= H_0+\sum_{k=1}^m u_k(t)H_k,
\end{equation}
in which $H_0$ and $H_k$'s are the free Hamiltonian and control Hamiltonians, respectively. The control fields $u_k(t)\in \mathbb{C}$, $k=1,\cdots,m$, are delivered to the quantum system via their respective control channels (e.g., control lines).

In most radio-frequency and microwave-based control systems, the control field generated by an arbitrary waveform generator (AWG) is prepared in piecewise-constant waveforms, i.e., \begin{equation}\label{}
  u_k(t)= u_k^j, \quad t_k^{j-1}\leq t\leq t_k^j,~~~j=0,1,\cdots,M,
\end{equation}
where $u_k^j$ and $t_k^j$ are the control amplitude and terminal time of the $j$th piecewise-constant control pulse of the $k$th field. Therefore, the control waveform is determined not only by the amplitude variables $\textbf{u}=\{u_k^j\}$, but also by the timing variables $\textbf{t}=\{t_k^j\}$.

When the pulses are periodically clocked and all control channels are perfectly synchronized, the timing variables ticks as $t_k^j=\bar{t}_k^j=jT_s$ for all $j=0,1,\cdots,M$ and $k=1,\cdots,m$, where $T_s$ is the sampling period of the AWG device. The timing errors are thus defined as the differences between the actual and ideal timing variables, i.e.,
\begin{equation}\label{}
\delta t_k^j=  t_k^j-\bar{t}_k^j = \tau_k +\xi_k^j,
\end{equation}
which consists of the latency time $\tau_k\geq0$ of the $k$th control channel and random jitter noises $\xi_k^j$ at each timing instant. The mean values of the jitter variables can be reasonably assumed to be zero.


\begin{figure}[htbp]
	\centering
	\includegraphics[width=1\columnwidth]{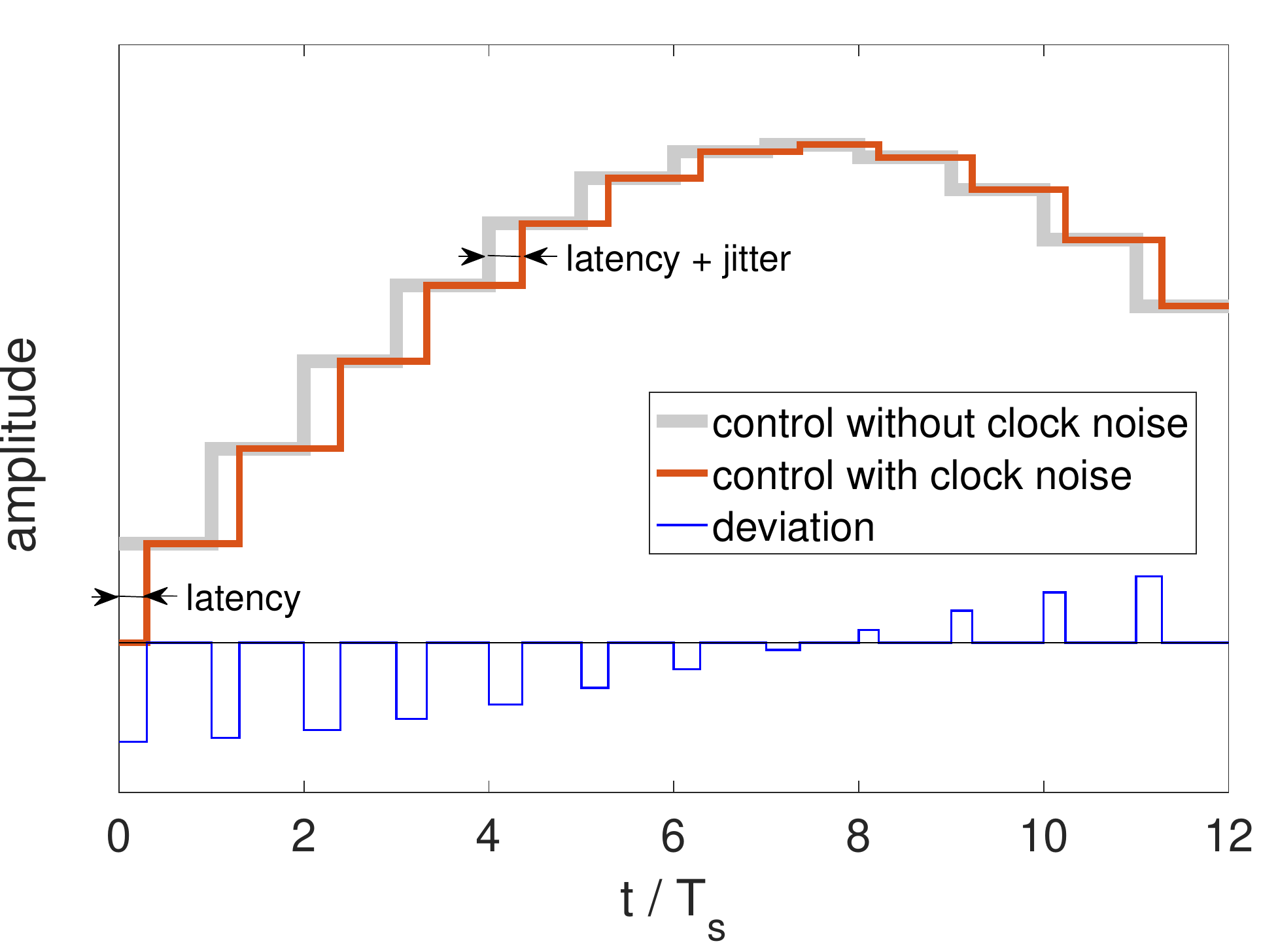}
	\caption{Schematic diagram for clock noises in piecewise-constant control signals. The (channel) latency is referred to as the arrival delay time of the AWG signal and the timing error of each sub-pulse consists of latency and random jitter noises. The deviation of the actual signal with clock noise from the ideally clocked signal consists of small sub-pulses described by Eq.~(\ref{eq:deltaut}).}
	\label{fig:ulatency}
\end{figure}

\subsection{Error analysis}
Throughout this paper, we assume that the amplitude variables ${\bf u}$ are precise (robust control to amplitude errors has been broadly studied in the literature, e.g., in~\cite{Vitanov2017}), and focus on the error caused by clock noises.

The clock noises $\delta t_k^j$ bring control errors into the quantum gate operation via the change of areas of the sub-pulses. Let
\begin{equation}\label{}
  \bar{u}_k(t)= u_k^j, \quad \bar{t}_{j-1}\leq t\leq \bar{t}_j,~~~j=0,1,\cdots,M,
\end{equation}
be the ideally clocked control signal and $\bar{U}(t)$ the unitary propagator steered by $\bar{u}(t)$, i.e.,
\begin{equation}\label{eq:ideal system}
i\dot{\bar{U}}(t)=\bar{H}(t)\bar{U}(t)=\left[H_0+\sum_{k=1}^m\bar{u}_k(t)H_k\right]\bar{U}(t).
\end{equation}
Then, the gate error operator $\Delta(t) = U(t)-\bar{U}(t)$ can be shown to obey the dynamical equation:
\begin{equation}\label{con:Edot}
i\dot{\Delta}(t)=  H(t)\Delta(t)+\sum_{k=1}^m \delta u_k(t) H_k \bar{U}(t),
\end{equation}
where $\Delta(0)=0$ and $\delta u_k(t)=u_k(t)-\bar{u}_k(t)$ is the difference between the actual and the ideally clocked control field in the $k$th channel. As can be seen in Fig.~\ref{fig:ulatency}, $\delta u_k(t)$ consists of the following slices of pulses:
\begin{equation}\label{eq:deltaut}
  \delta u_k(t) = \left\{
  \begin{array}{ll}
    \delta u_k^j, & \bar{t}_k^j<t<t_k^j \\
    0, & {\rm else}
  \end{array}\right.
\end{equation}
where $\delta u_k^j=u_k^{j-1}-u_k^{j}$ is the decremental control amplitude. One can immediately see that the gate error can be reduced by making $\delta u_k^j$ as small as possible, i.e., choosing slowly varying smooth control fields.

For a more accurate analysis, we can formally integrate Eq.~(\ref{con:Edot}) to evaluate the gate error at the final time $t=T$, as follows:
\begin{eqnarray}\label{con:E}
\Delta(T) &=& -iU(T)\sum_{k=1}^m\int_{0}^T \delta u_k(\tau) U^{\dag}(\tau)H_k \bar{U}(\tau){\rm d}\tau \nonumber \\
& = & -iU(T)\sum_{k=1}^m\sum_{j=0}^M \int_{\bar{t}_k^j}^{t_k^j} \delta u_k^j U^{\dag}(\tau)H_k \bar{U}(\tau){\rm d}\tau.
\end{eqnarray}
In the perturbation regime, i.e., when all timing errors $\delta t_k^j$ are much smaller than the sampling period $T_s$, it is reasonable to take the following approximation
\begin{equation}\label{}
  U(\tau)\approx \bar{U}(\tau)\approx \bar{U}(\bar{t}_k^j), \quad \forall \tau\in[\bar{t}_k^{j},{t}_k^{j}],
\end{equation}
which leads to a simplified formula:
\begin{equation}\label{}
\Delta(T)\approx -iU(T)\sum_{k=1}^m\sum_{j=1}^M \delta t_k^j \delta u_k^j \bar{H}_k^j,
\end{equation}
where $\bar{H}_k^j\triangleq \bar{U}^\dag(\bar{t}_k^j)H_k\bar{U}(\bar{t}_k^j)$ is only dependent on the ideal system (\ref{eq:ideal system}) but not on the clock noise. The Frobenius norm of $\Delta(T)$ can thus be expanded as
\begin{equation}\label{}
\|\Delta(T)\|_F^2\approx\sum_{k,k'}\sum_{j,j'} \delta t_k^j\delta t_{k'}^{j'} \delta u_k^j \delta u_{k'}^{j'}\cdot {\rm tr}\!\left(\bar{H}_k^j \bar{H}_{k'}^{j'}\right).
\end{equation}

Now we can quantify the robustness (or the sensitivity) of the control by the average gate error over the clock noises:
\begin{eqnarray}
&&J_{\rm N}[{\bf u}]=\langle\|\Delta(T)\|_F^2\rangle \nonumber \\
&\approx & \sum_{k,k'}\sum_{j,j'} \langle\delta t_k^j\delta t_{k'}^{j'}\rangle \delta u_k^j \delta u_{k'}^{j'} {\rm tr}(\bar{H}_k^j \bar{H}_{k'}^{j'}) \nonumber \\
&=& \sum_{k,k'}\!C^\tau_{kk'}{\rm tr}\!\left(\delta\textbf{H}_k \delta\textbf{H}_{k'}\right) + \mu_0^2\! \sum_{k}\!\|\delta\textbf{u}_k\|^2\|H_k\|^2_F, \label{eq:DeltaT}
\end{eqnarray}
where
\begin{equation}\label{}
\|\delta \textbf{u}\|^2= \sum_{j=1}^M|\delta u_k^j|^2,\quad  \delta \textbf{H}_k = \sum_{j=1}^{M}\delta u_k^j\bar{H}_k^j.
\end{equation}
and $C^\tau_{kk'}=\langle\tau_k\tau_{k'}\rangle$ is the covariance between the $k$th and $k'$th channel latencies $\tau_k$ and $\tau_{k'}$, and $\mu_0^2$ is the variance of the jitter errors (assumed to follow identical distribution with zero mean).

The estimation formula (\ref{eq:DeltaT}) indicates the respective roles of the latency and jitter noises. The latency affects the average error through the first term, a complicated function of the unitary propagators, while the second term shows that the jitter noises is determined by the smoothness of the control fields characterized by $ \|\delta \textbf{u}\|^2$.

\section{Robust Optimization algorithms for Clock noises}  \label{Sec:Three}
The task of robust control here is to find proper control amplitude variables $\textbf{u}=\{u_k^j\}$ that is as insensitive as possible to the clock noises ${\bf \delta t}=\{\delta t_k^j\}$ in the timing variables ${\bf t}=\{t_k^j\}$. In the following, we will propose two types of algorithms, one deterministic and one stochastic, for the design of robust and high-precision controls based on the average gate error analysis.

\subsection{Deterministic homotopic algorithm}
This algorithm is based on the error estimation formula (\ref{eq:DeltaT}) that quantifies the control robustness. Recall that a standard gradient algorithm (without consideration of robustness) minimizes $J_0[\textbf{u}]=\|\bar U(T)-U_{f}\|^{2}$ based on the ideal model. To improve the control robustness, it is natural to introduce (\ref{eq:DeltaT}) as a penalty term to the gate error objective as follows:
\begin{eqnarray}
J[\textbf{u}] &=& J_0[\textbf{u}]+\beta J_{\rm N}[\textbf{u}],\label{eq:GRAPE}
\end{eqnarray}
where $\beta>0$ is the weight parameter. The composite objective is deterministically dependent on the control variable ${\bf u}$, in which the clock noise variables have been averaged out. Thus, it can be minimized by a gradient-descent algorithm. However, the parameter $\beta$ must be be carefully chosen for balancing the objectives $J_0$ and $J_{\rm N}$. Otherwise, either $J_0$ or $J_{\rm N}$ may be sacrificed when they conflict with each other.

To avoid the unwanted trade-off between $J_0$ and $J_{\rm N}$, we propose a two-stage homotopic algorithm that minimizes the two objectives separatively. In the first stage, we apply the well-known gradient-based GRAPE algorithm \cite{Khaneja2005Optimal} to minimize the gate error starting from a randomly chosen initial guess on the field. This can be done, for example, by following the steepest gradient-descent direction of $J_0$:
\begin{equation}
\textbf{u}^{(\ell+1)} = \textbf{u}^{(\ell)} - \alpha_\ell \frac{\partial J_0}{\partial \textbf{u}^{(\ell)}},
\end{equation}
where $\textbf{u}^{(\ell)}$ are the control amplitudes in the $\ell$th iteration and $\alpha_\ell$ is the corresponding learning rate. The obtained optimal control can achieve extremely high precision, but its robustness to clock noises is not guaranteed.

In the second stage, we apply a homotopic algorithm \cite{Rothman2005,Rothman2006} that continuously reduces the value of $J_{\rm N}$ while maintaining the achieved high precision $J_0$. This is done by updating the control along a descending direction of $J_{\rm N}$ that is orthogonal to the gradient of $J_0$, which can be obtained by the following Schmidt orthogonalization:
\begin{equation}\label{eq:Homotopic}
\textbf{u}^{(\ell+1)} = \textbf{u}^{(\ell)} - \alpha_\ell \left[\frac{\partial J_{\rm N}}{\partial \textbf{u}^{(\ell)}}-\frac{\left(\frac{\partial J_{\rm 0}}{\partial \textbf{u}^{(\ell)}}\right)^T\frac{\partial J_{\rm N}}{\partial \textbf{u}^{(\ell)}}}{\left(\frac{\partial J_{\rm 0}}{\partial \textbf{u}^{(\ell)}}\right)^T\frac{\partial J_{\rm 0}}{\partial \textbf{u}^{(\ell)}}}\frac{\partial J_{\rm 0}}{\partial \textbf{u}^{(\ell)}}\right].
\end{equation}

The homotopic algorithm can be efficiently carried out owing to its deterministic nature, and both the control precision (through $J_0$) and the robustness (through $J_{\rm N}$) can be improved without having to make a compromise. Since the algorithm is based on the perturbation analysis of the clock-noise induced gate errors, it is supposed to be effective when the noise level is not high. Beyond the perturbation regime, one can consider the stochastic optimization algorithm to be proposed below.

\subsection{Randomized b-GRAPE algorithm}
To further deal with larger clock noises beyond the perturbation regime, we introduce our recently proposed b-GRAPE algorithm, a stochastic gradient-descent algorithm \cite{Wu2018}, for enhancing the control robustness against clock noises.

The basic idea is, instead of evaluating the average gate error with Eq.~(\ref{DeltaT}), to directly average the error over a set of randomly chosen samples of clock noises $\mathcal{S}=\{\delta\mathbf{t}_1,\delta\mathbf{t}_2,\cdots,\delta\mathbf{t}_K\}$ according to some {\it a priori} probability distribution, as follows
\begin{equation} \label{eq:bGRAPE}
J_\mathcal{S}[\textbf{u}]=\frac{1}{N}\sum_{{\bf {t}}\in\mathcal{S}}\|U(T,{\bf {t}})-U_f\|^2.
\end{equation}
$J_\mathcal{S}$ is a good measure on the average error as long as the sample set $\mathcal{S}$ is sufficiently large, and it does not matter whether the noise is in perturbation regime or not. Robust and high precision control can be achieved if $J_{\mathcal{S}}$ can be made sufficiently small. However, such sampling-based algorithm \cite{dong2015sampling} is computationally expensive when $K$ is very large, because the unitary evolution must be calculated for $N$ times corresponding to all samples in the set.

The b-GRAPE algorithm exploits the noise samples in a different way. It randomly picks a batch of samples for each iteration, say $\mathcal{S}^{(j)}=\{\textbf{t}_1^{(\ell)},\cdots,\textbf{t}_B^{(\ell)}\}$, and updates the control along the gradient evaluated with this batch, i.e.,
\begin{equation}
\textbf{u}^{(\ell+1)} = \textbf{u}^{(\ell)} - \alpha_\ell \frac{\partial J_{\mathcal{S}^{(\ell)}}}{\partial \textbf{u}^{(\ell)}},
\end{equation}
where $\alpha_\ell$ is the learning rate. Since the batch size does not have to be large, the b-GRAPE algorithm can be efficiently carried out (e.g., using parallel computing), though the iteration process is noisy due to the randomly sampled batches. More importantly, b-GRAPE optimization using small batches can more easily find highly robust solutionsbecause the large artificial noises tend to steer the search away from poor non-robust solutions.

\section{Simulation Results} \label{sec:Simulation}
In this section, we test the proposed robust control algorithms with an example of two-qubit control system, whose Hamiltonian is as follows:
\begin{equation}
H(t)=g\sigma_{z1}\sigma_{z_2}+\sum_{k=1}^{2}[u_{k}(t)\sigma_{k}^++u_{k}^*(t)\sigma_{k}^-],
\end{equation}
where $\sigma_{zk}$ ($k=1,2$) is the Pauli matrix of the $k$th qubit and the coupling strength is $g = 2\pi \times 10$MHz. We aim to implementing a CNOT gate that is robust to clock noises. This system involves four independent control functions (including the real and imaginary parts of $u_{k}(t)=u_{kx}(t)+iu_{ky}(t)$, $k = 1,2$). They are delivered to the qubits through two control channels (e.g., two separate control lines in a superconducting quantum computing system). Therefore, there are two mutually independent clock noises in this system.

\begin{figure}[htbp]
	\centering
   \includegraphics[width=1\columnwidth]{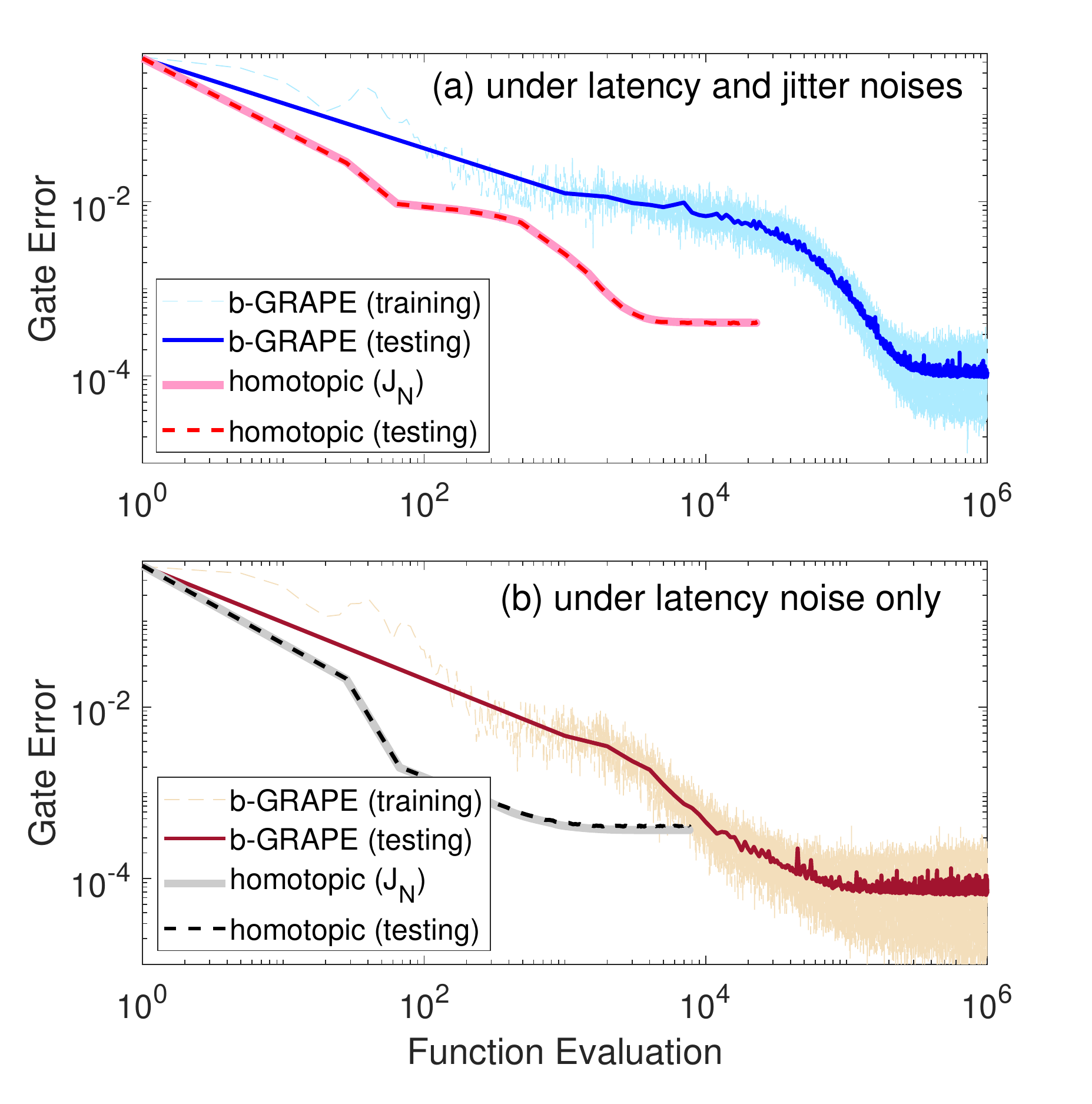}
	\caption{The learning curves of homotopic and b-GRAPE algorithms under (a) latency and jitter noises and (b) under latency noise only. The average gate errors estimated in the homotopic algorithm and in the training process of b-GRAPE algorithm are both consistent with their testing performances (each intermediate control field is tested by $10^4$ random clock noise samples).} 
	\label{fig:learning curve}
\end{figure}

In the simulation, the sampling period is chosen to be $T_s=1$ns and the pulse length be $T=50$ns (i.e., $M=50$ amplitude variables are to be optimized). The latencies $\tau_1$ and $\tau_2$ in the two channels are both uniformly sampled between $0$ns and $0.4$ns (i.e., $0-40\%$ of the sampling period). The covariance matrix can be calculated to be
\begin{equation}\label{}
  C^\tau=\left(
           \begin{array}{cccc}
    0.0533 &   0.0533 &   0.0400 &   0.0400 \\
    0.0533 &   0.0533 &   0.0400 &   0.0400 \\
    0.0400 &   0.0400 &   0.0533 &   0.0533 \\
    0.0400 &   0.0400 &   0.0533 &   0.0533
            \end{array}
         \right){\rm ns}^2,
\end{equation}
where each clock noise affects two controls in the corresponding channel. The jitter noises $\xi_k^j$ are uniformly sampled between $-0.05$ns and $0.05$ns (i.e., $\pm 5\%$ of the sampling period), whose variance is $\mu_0^2=8.33\times 10^{-4}{\rm ns}^2$. For comparison, we also perform the optimization with only latency noise (i.e., with the same $C^\tau$ but $\mu_0^2=0$). The simulation results of these two cases are shown in Fig.~\ref{fig:learning curve}, in which both the homotopic and the b-GRAPE algorithms are applied.

In the homotopic algorithm, we first use a standard GRAPE algorithm to minimize the gate error $J_0$ down to machine precision based on the ideal control model (\ref{eq:ideal system}). Then, we follow the projected gradient (\ref{eq:Homotopic}), in which the above calculated statistical parameters $\mu_0^2$ and $C^{\tau}$ are involved, to reduce the average gate error $J_{\rm N}$ while keeping $J_0$ to be sufficiently small. When $J_0$ rises up (above $10^{-6}$) due to the numerical error, we apply the GRAPE algorithm to correct it back to below $10^{-10}$. The actual robustness of  each intermediate control field is evaluated by the average error tested over $10^4$ random samples. From the simulations, the statistic average gate error using testing samples agrees very well with $J_{\rm N}$ calculated by the estimation formula (\ref{eq:DeltaT}), showing that $J_{\rm N}$ can be used as a good approximation on the average gate error, and hence the robustness measure of control fields against clock noises.

As for the b-GRAPE algorithm, the batch size is chosen to be $B=5$ and the clock noises are sampled during the optimization according to the same probability distribution described above. The training curves are very noisy due to the chosen small batch size, but the testing curve (i.e., evaluating the average error for each intermediate control field with $10^4$ additional random noise samples) exhibits a steady decrease of the gate error.

\begin{figure}[htbp]
	\centering
    \includegraphics[width=1\columnwidth]{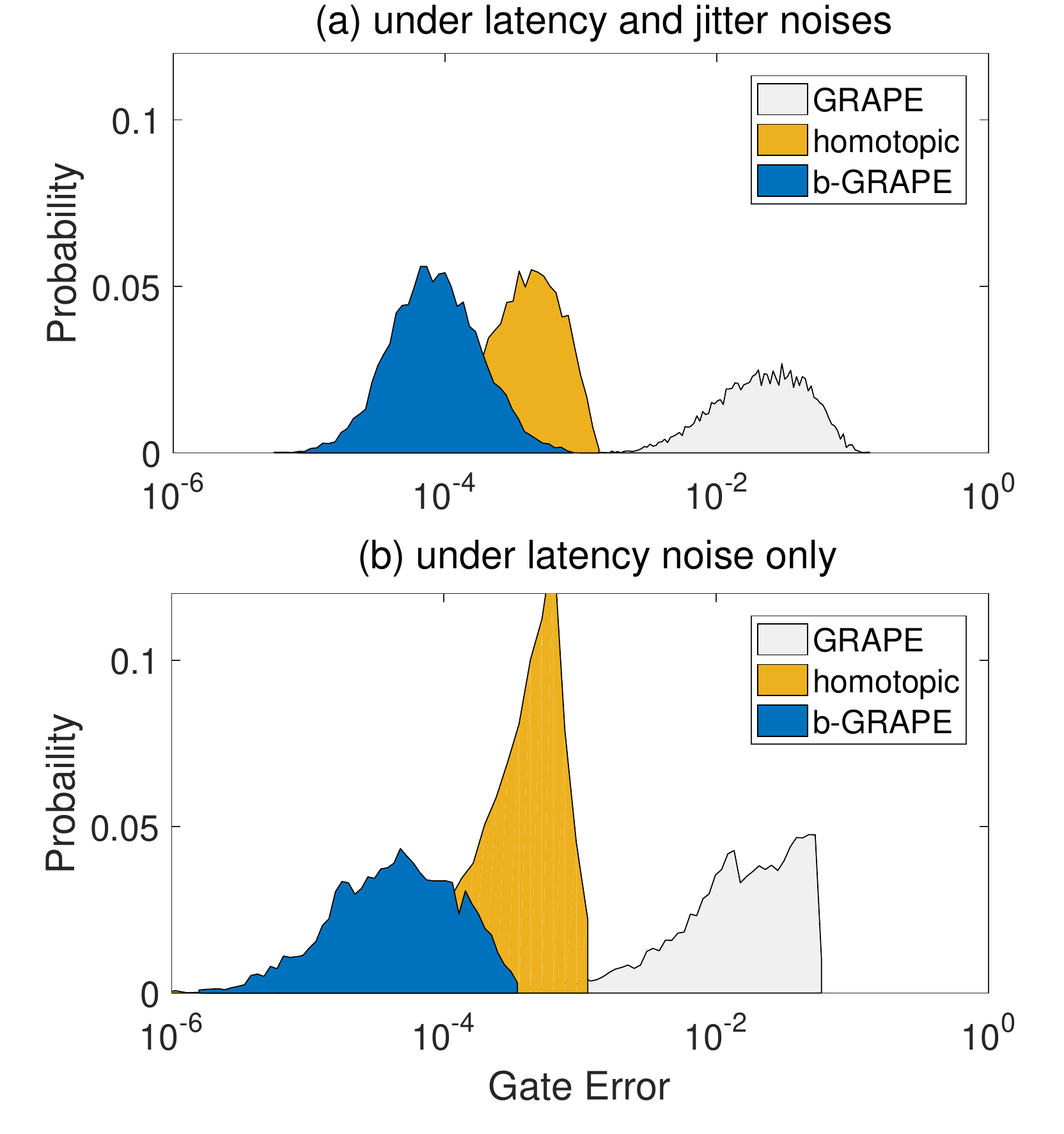}
	\caption{The two-qubit gate (CNOT) error distribution under optimal controls obtained by GRAPE, homotopic and b-GRAPE algorithms (a) under latency and jitter clock noises and(b) under latency noise only. The robust optimization can effectively push the entire error distribution to the high-precision regime.}
	\label{fig:error distribution}
\end{figure}

The simulation results show that both algorithms can effectively enhance the control robustness against clock noises. The homotopic algorithm converges faster, but the ultimate average gate error it achieved is larger than that achieved by b-GRAPE algorithm. The difference is also shown in the analysis of gate error distribution in Fig.~\ref{fig:error distribution}. In addition,  both the distributions corresponding to the homotopic and GRAPE algorithms are skewed to the large-error side, while that corresponding to the b-GRAPE algorithm is almost normal. Therefore, the b-GRAPE algorithm tends to find more robust control fields.

\begin{figure}[htbp]
	\centering
    \includegraphics[width=1\columnwidth]{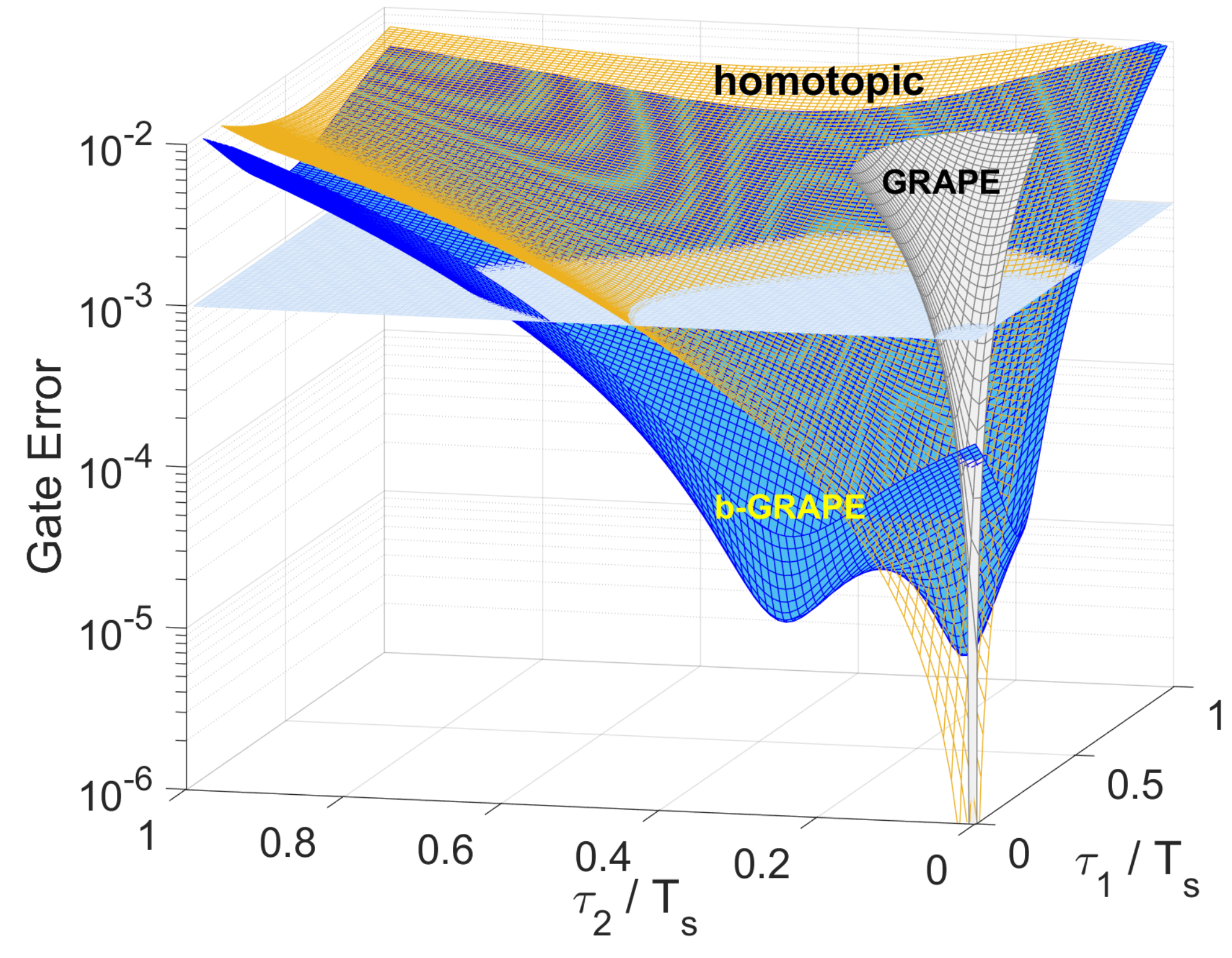}
	\caption{The dependence of gate error on two latency variables under optimal controls obtained by GRAPE, homotopic and b-GRAPE algorithms. }
	\label{fig:3Dplot}
\end{figure}

Figure \ref{fig:3Dplot} shows how the gate error relies on the two latency noise parameters $\tau_1$ and $\tau_2$, where the jitter noise is absent. The fields obtained by b-GRAPE and homotopic algorithms can suppress the error below $10^{-3}$ over a large region, but GRAPE (without consideration of robustness) maintains high-precision only in a much smaller region. The homotopic algorithm is designed to always maintain highest control precision at $\tau_1=\tau_2=0$ns (corresponding to the noiseless ideal system). Interestingly, to achieve the overall high control precision, b-GRAPE algorithm adapts itself better to the noise distribution, as its highest precision at $\tau_1=0.24$ns and $\tau_2=0.06$ns is closer to the center of the sampling region. This explains why b-GRAPE algorithm can find more robust controls.

\begin{figure}[htbp]
	\centering
   \includegraphics[width=1\columnwidth]{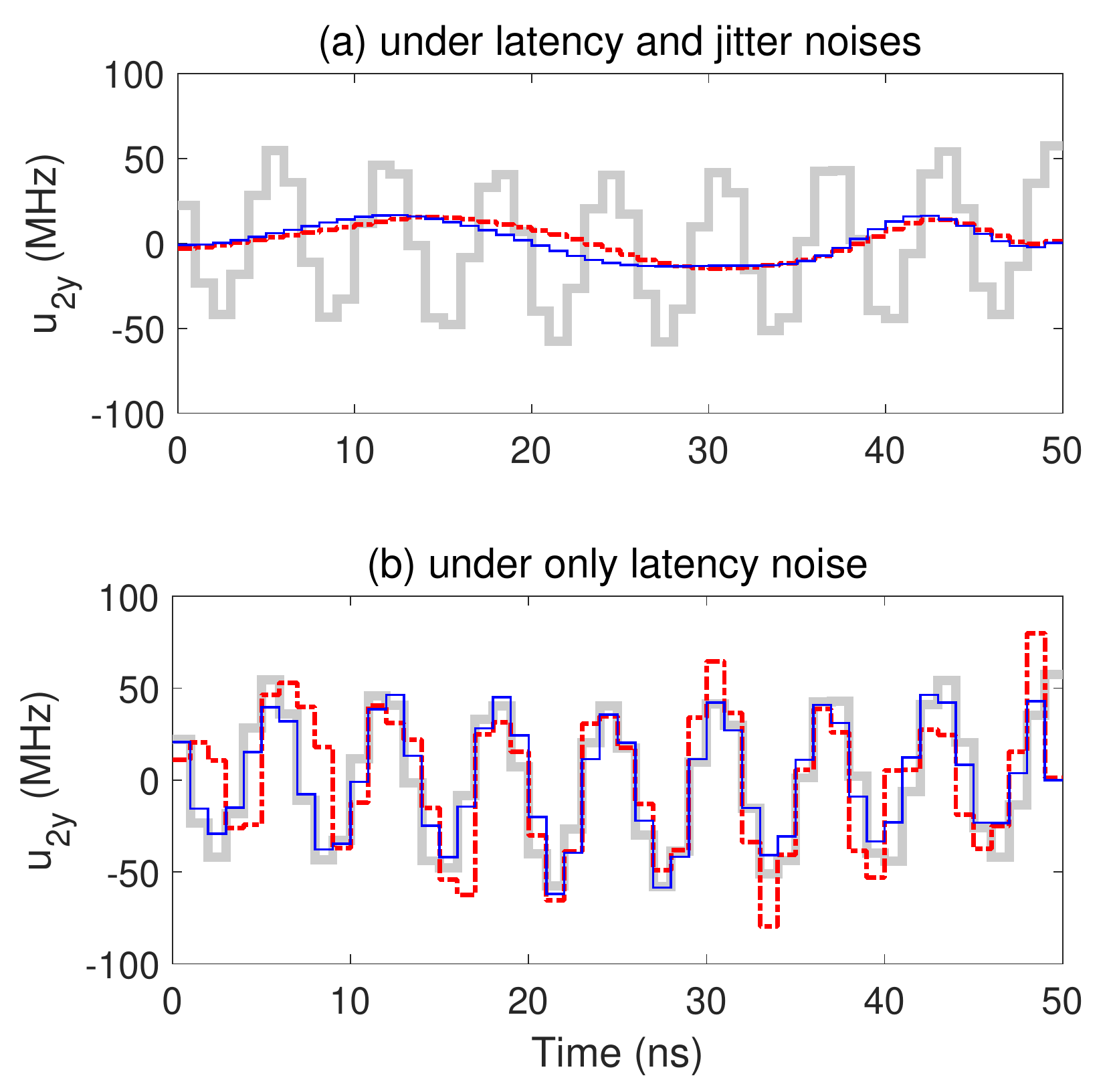}
	\caption{The optimized $y$-axis control fields on the second qubit obtained by GRAPE (grey solid curve), homotopic (blue solid curve) and b-GRAPE (red dash-dotted curve) algorithms. The robust optimization against random jitter noises leads to much smoother waveforms shown in (a) than without jitter noises in (b).}
	\label{fig:fields}
\end{figure}

We also plot the optimized fields in Fig.~\ref{fig:fields}. Taking the $y$-axis field $u_{2y}(t)$ on the second qubit as an example, we can see that the waveforms optimized with or without jitter noises are very different. The presence of jitter noise leads to much smoother fields because it penalizes the average error through the smoothness term $\|\delta {\bf u}\|^2$ derived in Eq.~(\ref{eq:DeltaT}). By contrast (see Fig.~\ref{fig:fields}b), the smoothness is almost unchanged when training the controls without jitter noises.

\section{Conclusion} \label{Sec:Con}
To conclude, we have analyzed the robustness of quantum control against clock noises that are prevalent in multi-channel control signal generation devices. Two algorithms are introduced for improving the robustness by minimizing the average gate error induced by clock noises, and their effectiveness on robustness improvement is shown by simulation examples. In comparison, the homotopic algorithm converges faster but the b-GRAPE algorithm can achieve higher performance owing to its better adaptivity to the noise.

The proposed algorithms represent two frameworks, deterministic and stochastic, for optimizing robust control against clock noises. More advanced optimization techniques can be incorporated to further improve the control robustness and precision, e.g., Newton-Ralphson algorithms for faster convergence or momentum-based skills for stabilizing of training during b-GRAPE optimization.

Realistic quantum systems always involve multiple noises or uncertainties in addition to clock noises. Whether it is possible, and how to design controls when these noises coexist, is much more challenging. These problems are to be explored in future.


\end{document}